\documentclass[Afour,sagev,times]{sagej}

\usepackage{moreverb,url}

\usepackage[colorlinks,bookmarksopen,bookmarksnumbered,citecolor=red,urlcolor=red]{hyperref}

\usepackage{tabulary,xcolor}
\usepackage{amsfonts,amsmath,amssymb}
\usepackage[T1]{fontenc}

\usepackage{hyperref}

\usepackage{ifluatex}

\usepackage{float}
\usepackage{amsfonts}
\usepackage{dsfont}
\usepackage{graphicx}
\usepackage{wrapfig}
\usepackage{lipsum}
\usepackage{amsmath}

\usepackage{amsfonts}
\usepackage{dsfont}

\usepackage{url,multirow,morefloats,floatflt,cancel,tfrupee}

\usepackage{colortbl}
\usepackage{xcolor}
\usepackage{pifont}
\usepackage[nointegrals]{wasysym}

\usepackage{ifxetex}

\usepackage{algorithm}
\usepackage[algo2e,linesnumbered,ruled,vlined]{algorithm2e}
\usepackage{nomencl}
\usepackage{algpseudocode}

\newtheorem{definition}{Definition}[section]

\newtheorem{theorem}{Theorem}[section]

\newcommand{\kw}[1]{{\color{blue}\textsf#1}}

\newcommand{\holl}{\text{HOL4}}

\newcommand\BibTeX{{\rmfamily B\kern-.05em \textsc{i\kern-.025em b}\kern-.08em
T\kern-.1667em\lower.7ex\hbox{E}\kern-.125emX}}

\def\volumeyear{2016}

\begin{document}

\runninghead{S. Murtza et al.}

\title{Towards the Formal Performance Analysis of Multistate Coherent Systems using HOL Theorem Proving}

\author{Shahid Ali Murtza, Waqar Ahmed, Adnan Rashid and Osman Hasan}

\affiliation{{School of Electrical Engineering and Computer Science (SEECS)} \\
{National University of Sciences and Technology (NUST), Islamabad, Pakistan}
}

\corrauth{Adnan Rashid, {School of Electrical Engineering and Computer Science (SEECS)} \\
{National University of Sciences and Technology (NUST), Islamabad, Pakistan}}

\email{adnan.rashid@seecs.nust.edu.pk}


\begin{abstract}
Many practical engineering systems and their components have multiple performance levels and failure modes. If these systems form a monotonically increasing structure function (system model) with respect to the performance of their components and also if all of their components affect the overall system performance, then they are said to be multistate coherent systems. Traditionally, the reliability analysis of these multistate coherent systems has been carried out using paper-and-pencil or simulation based methods. The former method is often prone to human errors, while the latter requires high computational resources for large and complex systems having components with multiple operational states. As a complimentary approach, we propose to use Higher-order-logic (HOL) theorem proving to develop a sound reasoning framework to analyze the reliability of multistate coherent systems in this paper. This framework allows us to formally verify generic mathematical properties about multistate coherent systems with an arbitrary number of components and their states. Particularly, we present the HOL formalization of series and parallel multistate coherent systems and formally verify their deterministic and probabilistic properties using the \holl~theorem prover. For illustration purposes, we present the formal reliability
analysis of the multistate oil and gas pipeline to demonstrate the effectiveness of our proposed framework.
\end{abstract}

\keywords{Multistate Coherent Systems, Reliability Analysis, Probabilistic Analysis, Theorem Proving, Higher-order Logic, \holl}

\maketitle


\section{Introduction}
In many real-life scenarios, engineering systems and their components assume an entire range of different performance levels or states varying from perfect functioning to complete failure. Such types of systems are known as multi-state systems~\citep{el1978multistate}. If a multistate system forms a monotonically increasing structure function (system model) with respect to the performance of its components and the overall system performance is dependent on each one of its component, then it is said to be a multistate coherent system~\citep{el1978multistate}. The monotonicity property of a multistate coherent system is described as each system component contributes to the system performance in a manner that if it fails, then the system performance also degrades or remains the same but does not improve or vice versa~\citep{ebrahimi1984multistate}. Similarly, a component is said to be relevant to the system performance, if it has a direct impact on some performance levels~\citep{ebrahimi1984multistate}. To analyze the performance of these complex multistate coherent systems, it is often required to evaluate the reliability of these systems at all possible performance levels (states) in terms of all the possible operational states of their components.

Traditionally, simulation based reliability analysis tools, such as ReliaSoft~\citep{coroporation2009reliability} and FReET~\citep{menvcik2016software}, have been used to analyze the reliability of multistate coherent systems. These simulation tools mainly utilize Monte-Carlo simulation, which is based on numerical approximation algorithms~\citep{billinton1991hybrid,zio2007monte}. Due to the large number of states involved in analyzing the reliability of complex multistate coherent systems, these simulation tools are computationally expensive and can only analyze very small sized systems. In contrast, the analytical approaches are more scalable than simulation based methods. For instance, El-Neweihi et al.~\citep{el1978multistate} developed an axiomatic theory to analytically analyze the performance of multistate coherent systems. They described that the performance of multistate coherent system is lower bounded by the performance of a series system and upper bounded by a parallel system. They presented several key deterministic and probabilistic properties, such as system concatenation (redundancy), and critical upper connection vector bounds, about the multistate coherent systems.
However, the manual approach of using these properties for conducting the performance analysis of real-world multistate safety-critical systems is impractical as well as prone to human error since these systems involve a large number of operational states.

Formal methods~\citep{hasan2015formal} are computer-based mathematical analysis techniques that involve constructing a mathematical model based on an appropriate logic and verification of its various properties based on deductive reasoning. Higher-order-logic (HOL) theorem proving is a sound reasoning environment and is widely adopted formal method for performing an accurate analysis of the engineering systems.
It has also been used for developing a formal dependability analysis framework~\citep{ahmad2017formal}, based on Reliability Block Diagram (RBD)~\citep{trivedi1982probability} and Fault Tree (FT)~\citep{trivedi1982probability} based modeling techniques, which have further been used to perform the reliability analysis of oil and gas pipelines~\citep{ahmad2018formal} and a railway traction drive system~\cite{ahmad2019formalization}, failure analysis of satellite solar arrays~\citep{ahmad2015towards} and an air traffic management system~\citep{ahmad2016formalization}. However, this framework~\citep{ahmad2017formal} is only limited to the binary state systems and cannot be used to analyze the performance analysis of complex multistate coherent systems.

\begin{figure*}[!ht]
	\centering
	\includegraphics[width=0.8\linewidth,scale=0.45]{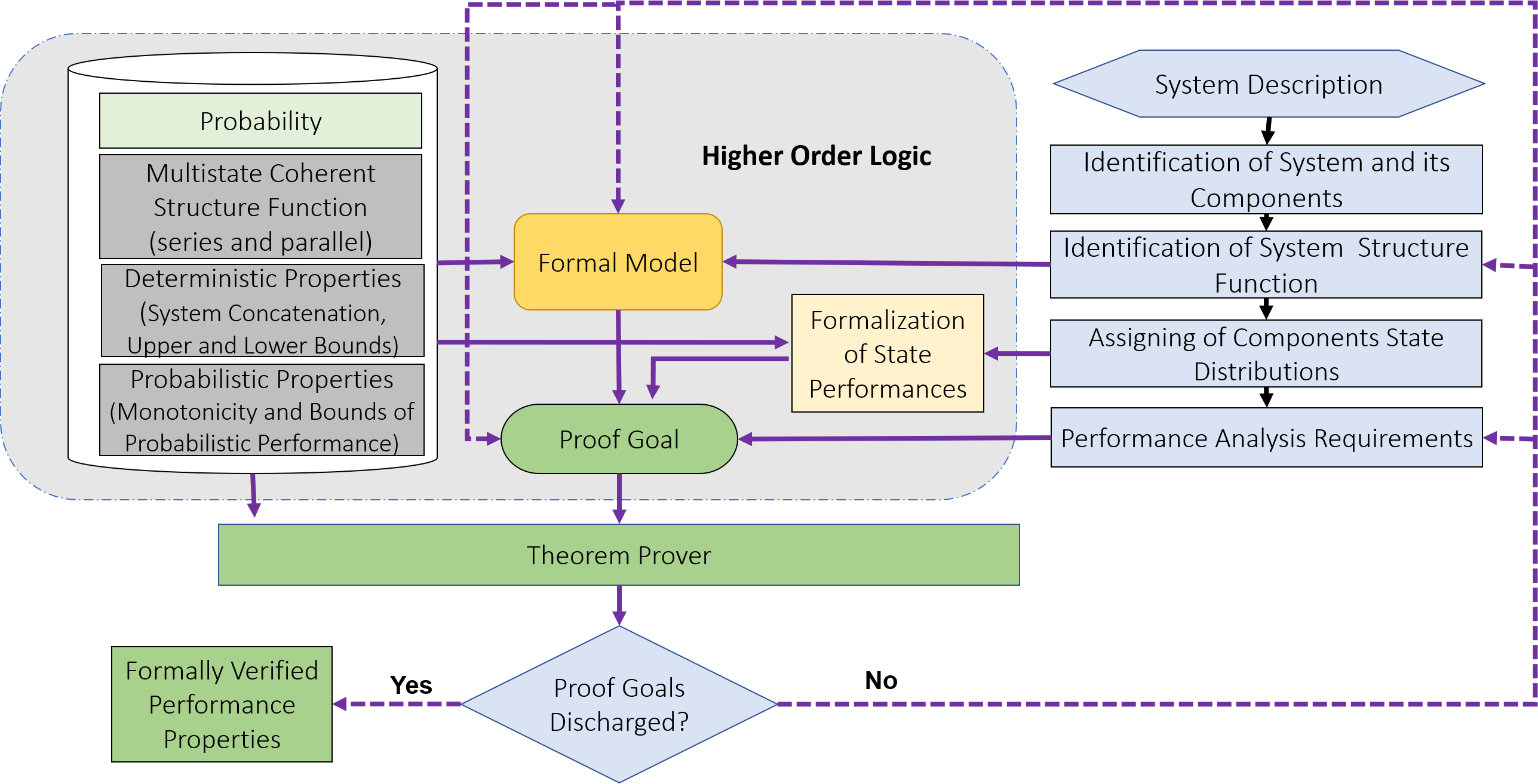}
	\caption{Proposed Framework}
	\label{Fig:prop_framework}
\end{figure*}

In this paper, we propose to use HOL theorem proving, as shown in Figure~\ref{Fig:prop_framework}, to conduct the performance analysis of multistate coherent systems. We particularly formalize the notion of multistate coherent systems and formally verify their deterministic and probabilistic properties, such as upper and lower bounds, system concatenation (redundancy), and critical upper connection vector bounds. To illustrate the practical effectiveness of our proposed approach, we conduct the formal performance analysis of the multistate oil and gas pipeline, which is modeled as a series and parallel multistate coherent system.

The rest of the paper is organized as follows: Section~\ref{SEC:related_work} provides a brief description of various analysis techniques used for performing the probabilistic performance analysis of systems.
We present preliminaries that include a brief introduction of theorem proving, the \holl~theorem prover and the formalization of probability theory in \holl, in Section~\ref{SEC:prelim}.
Section~\ref{SEC:multistate_coherent_systems} provides our formalization of multistate coherent systems, i.e., formalization of the multistate series and parallel structure functions and formal verification of their deterministic properties. Section~\ref{SEC:stoc_perf_multistate_coherent_systems} presents the stochastic performance of the multistate coherent systems. We provide the formal performance analysis of multistate oil and gas pipeline system in Section~\ref{SEC:case_study}. Finally, Section~\ref{SEC:conclusions} concludes the paper.


\section{Related Work} \label{SEC:related_work}

In this section, we present a brief overview of different analysis techniques that have been used for the probabilistic performance analysis of the multistate coherent systems.


\subsection{Simulation} \label{SUBSEC:simulation}

Many Monte-Carlo based simulation approaches have also been used for assessing the reliability of multistate coherent systems. For instance, Billinton et al.~\citep{billinton1991hybrid} proposed a hybrid approach by combining Monte-Carlo simulation and an enumeration technique. They also analyzed large scale composite generation transmission systems by modeling its generation units as multistate system. Similarly, Zio et al.~\citep{zio2007monte} proposed a Monte Carlo simulation based methodology in order to model the complex dynamics of multi-state components by considering their operational dependencies in the overall system state. As an extension to their work, they also applied Monte-Carlo simulation based methodology to evaluate different reliability importance measures of the components at a given state in a multi-state series-parallel systems~\citep{zio2004estimation}. Jose et al.~\citep{ramirez2005monte} presented a Monte-Carlo simulation based approach to evaluate the reliability of a multistate system. They applied this methodology to a multi-state system involving two-terminal reliability computation. These simulation based methodologies pose a limit on their usability in terms of computational cost when the number of components or their performance levels become large.


\subsection{Analytical Methods} \label{SUBSEC:analytical_methods}

In the literature, many analytical approaches proposed novel methods to evaluate the reliability of multi-state systems. A simplest approach among them is the extension of existing binary models to multi-state cases. Caldarola~\citep{caldarola1980coherent} used a Boolean variable to each component state in a multistate system. This approach also presented a method to reduce a multistate system to a binary system. Boedidheimer et al.~\citep{boedigheimer1994customer} utilized the Binary Decision Diagram (BDD) approach to determine the reliability of multistate systems. Their methodology was basically to define a customer-centered structure function that requires the customers to assign an appropriate number of states for components and the entire system. Some existing literature also proposed the stochastic process based analytical methods for reliability evaluation of multistate systems. For instance, Natvig and Streller~\citep{natvig1984steady} were the first to introduce the stochastic process approach for reliability analysis of multistate systems. Similarly, Aven~\citep{aven1999stochastic} applied modern theory of stochastic process to present a more advanced probabilistic framework. This method greatly facilitated the users to develop general failure models in order to get formulas for evaluation of various performance measures while conducting the reliability analysis of multistate systems in complex situations. Xue et al.~\citep{xue1995dynamic} presented the idea of combining the Markov processes and coherent structure function for multistate systems.

El-Neweihi et al.~\citep{el1978multistate} developed an axiomatic theory to analytically analyze the performance of multistate coherent systems. This work utilizes the idea of structure function that describe the relation of system states to the states of its components. As compared to other analytical approaches in this domain, their work is simplistic and systematic representation of the multistate coherent systems and their deterministic and probabilistic properties.

All these analytical approaches are carried out manually using paper-and-pencil based methods and thus are prone to human errors when the system is large and complex.


\subsection{Proposed Approach} \label{SUBSEC:proposed_approach}

Recently, a formal dependability analysis framework~\citep{ahmad2017formal}, based on Reliability Block Diagram (RBD)~\citep{trivedi1982probability} and Fault Tree (FT)~\citep{trivedi1982probability} modeling techniques, has been developed using HOL theorem prover. This framework has been successfully utilized to carry out the reliability analysis of oil and gas pipelines~\citep{ahmad2018formal} and a railway traction drive system~\citep{ahmad2019formalization}, failure analysis of satellite solar arrays~\citep{ahmad2015towards} and an air traffic management system~\citep{ahmad2016formalization}. Ahmad et al.~\citep{ahmad2019formalization} also presented the formalization of reliability importance measures and analyzed a railway signaling system. However, this framework~\citep{ahmad2017formal} is only limited to the binary state systems and cannot be used to analyze the performance analysis of complex multistate coherent systems. In this work, we formalize from scratch, the notion of multistate coherent systems and their probabilistic analysis using HOL theorem proving. The main idea is to analyze the performance of multistate coherent systems by considering the performance states of each of their components. We particularly formalize the work of El-Neweihi et al.~\citep{el1978multistate}, to build the formal reasoning support to conduct the reliability analysis of multistate coherent series and parallel systems. We also present the HOL formalization of their deterministic and probabilistic properties essentially required to analyze the system performance. To the best of our knowledge, this is the first formal work describing the formalization of the multistate coherent system using HOL theorem proving.


\section{Preliminaries} \label{SEC:prelim}
This section presents a brief introduction to theorem proving, the \holl~theorem prover and the formalization of probability theory in the \holl~theorem prover to facilitate the understanding of the rest of the paper.


\subsection{Theorem Proving} \label{SUBSEC:theorem_proving}
Theorem proving~\citep{gordon_89} is a broadly used formal verification technique. The system that we wish to analyze is first mathematically modeled in an appropriate logic and then its properties of interest are verified using computer-based formal tools. The use of formal logics as a modeling approach makes theorem proving a very adaptable verification tool as it is feasible to formally verify any system that can be presented mathematically. The core of theorem provers typically consists of some universal axioms and primitive inference rules. Soundness is guaranteed as each new theorem should be created from these fundamental or already proved theorems and primitive inference rules. The verification effort of a theorem in a theorem prover varies from trivial to complicated, relying on the underlying logic~\citep{harrison_96a}.


\subsection{\holl~Theorem Prover} \label{SUBSEC:hol_theorem_prover}

\holl~\citep{gordon1993introduction} is an interactive theorem prover that was developed at the University of Cambridge, UK for proving theorems in higher-order logic. This makes use of the simple type theory of Church~\citep{church_40} along with Hindley-Milner polymorphism~\citep{milner_77} to implement higher-order logic. It has been successfully employed as a verification methodology for both software and hardware as well as a platform for the formalization of pure mathematics. The \holl~core consists of only $4$ basic axioms and $8$ primitive inference rules, which are implemented as Meta Language (ML~\cite{paulson1996ml}, a general-purpose functional programming language) functions. The ML’s type system ensures that only valid theorems can be constructed. Soundness is assured as every new theorem must be verified by applying these basic axioms and primitive inference rules or any other previously verified theorems/inference rules. In this work, we utilize the \holl~theories of Boolean, lists, sets, positive integers, \emph{real} numbers, measure and probability~\citep{mhamdi_11}. In fact, one of the primary motivations of selecting the \holl~theorem prover for our work was to benefit from these built-in mathematical theories. Table~\ref{TAB:hol_symbols_n_functions} provides the mathematical interpretations of some frequently used \holl~symbols and functions, which are inherited from existing \holl~theories.

\begin{table*}[h]
	\centering
\renewcommand{\arraystretch}{0.65}
\begin{tabular}{|c | c | c|}
\hline
\textbf{\holl~Symbols} & \textbf{Standard Symbols} & \textbf{Meaning} \\
\hline
/\textbackslash & and & Logical \textit{and} \\ \hline
\textbackslash / & or & Logical \textit{or} \\ \hline
$\neg$ & not & Logical \textit{negation} \\ \hline
$\mathtt{::}$  &  $cons$ & Adds a new element to a list  \\ \hline
$\mathtt{MEM\ a\ L}$  &  $member$ & True if $a$ is a member of list $L$ \\ \hline
\texttt{LENGTH L} & \textit{length} & Length of list L  \\  \hline
\texttt{(a,b)} & \textit{a $\times$ b}& A pair of two elements \\ \hline
\texttt{FST} & \textit{fst(a,b) = a} & First component of a pair  \\  \hline
\texttt{SND} & \textit{snd(a,b) = b}& Second component of a pair  \\  \hline
\texttt{$\lambda$x.t} & \textit{$\lambda$x.t} & Function that maps \textit{x to t(x)} \\ \hline
\texttt{\{x|P(x)\}} & $\{\lambda x.P(x)\}$  & Set of all \textit{x} such that \textit{P(x)} \\ \hline
\texttt{SUC n} & \text{(n + 1)} & Successor of natural number \\ \hline
\end{tabular}
\caption{\holl~Symbols and Functions}
\label{TAB:hol_symbols_n_functions}
\end{table*}


\subsection{Probability Theory} \label{SUBSEC:prob_theory}

In mathematics, a measure space is described as a triple $(\Omega,\Sigma,\mu)$, where $\Omega$ is a sample space set, $\Sigma$ denotes a $\sigma$-algebra of subsets of $\Omega$, where the subsets are usually known as measurable sets, and $\mu$ is a measure on domain $\Sigma$. A probability space defined on a measure space $(\Omega,\Sigma,Pr)$, such that the measure, denoted by $Pr$, is called the probability with the measure of sample space $\Omega$ equal to $1$, i.e., $Pr(\Omega)=1$. In the \holl~formalization of probability theory~\cite{mhamdi_11}, given a probability space $p$, the functions \texttt{space}, \texttt{subsets} and \texttt{prob} return the corresponding $\Omega$, $\Sigma$ and $Pr$, respectively. Probability theory in \holl~provides many basic probability axioms which have been formally verified, that play an important role in formal reasoning about probabilistic properties of multi-state coherent systems. Random variable is the function defined on sample space. It takes all the possible outcome of sample space and maps them to mathematically easier outcome labels, usually a real number.


\section{Multistate Coherent Systems} \label{SEC:multistate_coherent_systems}

To facilitate the reader understanding, we first present the basic notations and the terminologies related to the mathematical representation of multistate systems. The notation $\mathbf{x} = (x_1,\cdots,x_n)$ denotes the state vector representing the operational states of components $1, \cdots, n$. $S=\{0,1,\cdots,M\}$ represents the set of operational states that a system and its component can acquire as performance levels.

\begin{itemize}
	\item[\textbullet] $(j_i, \mathbf{x}) \equiv (x_1,\cdots,x_{i-1},j,x_{i+1},\cdots,x_n)$, where $j= 0, 1, \cdots, M$, describes that value $j$ is substituted at $i^{th}$ index of vector $\mathbf{x}$.
	
	\item[\textbullet]  $(\mathbf{\cdot}_i, \mathbf{x}) \equiv (x_1,\cdots,x_{i-1},\mathbf{\cdot},x_{i+1},\cdots,x_n)$, where $\cdot$ describes that an arbitrary value is substituted at $i^{th}$ index of vector $\mathbf{x}$.
	
	\item[\textbullet] $\mathbf{j} \equiv  (j,j, \cdots,j)$ is a state vector $\mathbf{j}$ with all values equal to $j$.
	
	\item[\textbullet] $\mathbf{y} \wedge \mathbf{x} \equiv (y_1 \wedge x_1,\cdots,y_n \wedge x_n) $, where $y \wedge x = min(y,x)$, defines the relationship between the corresponding elements of two equal length vectors returning a vector with entries of the minimum value of the corresponding elements. The relation $y \wedge x = min(y,x)$ can be viewed as the components $x$ and $y$ that are connected in series.
	
	\item[\textbullet] 	$\mathbf{y} \vee \mathbf{x} \equiv (y_1 \vee x_1,\cdots,y_n \vee x_n)$, where $y \vee x = max(y,x)$, returns a vector with entries of the maximum value of the corresponding elements. The relation $y \vee x = max(y,x)$ can be viewed as the components $x$ and $y$ that are connected in parallel.
	
	\item[\textbullet] $\mathbf{y} < \mathbf{x}$, where $y_i \le x_i$ for $i = 1, \dots, n$, and $y_i < x_i$ for some $i$, ensures that all the elements (components states) in $\mathbf{y}$  are less than or equal to the corresponding elements in $\mathbf{x}$ or at least one element in $\mathbf{y}$  is less than the corresponding element in  $\mathbf{x}$.
	
	\item[\textbullet] Describing $f(x_1,\cdots,x_n)$  as an increasing function states that $f$ is increasing in each of its argument.
	\item[\textbullet] Given a distribution function $F$, the notation $\overline{F}$ represents its complement $1-F$ in the probability space.
\end{itemize}

Consider a system having $n$ components while both the system and its components can assume a set of distinct finite states
depicting different levels of performance varying from perfect functioning mode
(State $M$) to complete failure mode (State $0$). As the time of operation increases, the component performance, given it was working in perfect functioning state $M$ initially,  degrades and enters a State $M - 1$, and upon further degradation it enters the state
$M-2$. Finally, the component reaches the State $0$, whereas an identical sequence of
decreasing state levels illustrates the working of system over time.

The vector $\mathbf{x}=(x_1, x_2, \ldots, x_i, \ldots, x_n)$ represents  a state vector with $n$ components, where the performance level $x_i$ of component $i$ presumes a value in the set  $S = \{0, 1, \cdots, M\}$.  In this work, we consider that the performance of the system is completely determined by the performance of its components. Thus, the state of the system is related to the states of its components by a function $\phi(\mathbf{x}):S^n \rightarrow S$. In other words, if we know the state vector $\mathbf{x}$, then we can obtain the value of $\phi(\mathbf{x})$ representing the system state. The function $\phi(\mathbf{x})$ is also known as the \textit{structure function} of the system.

According to El-Neweihi et al.~\cite{el1978multistate}, a multistate system qualifies to be coherent if its structure function, $\phi(\mathbf{x})$, meets the following conditions:

\begin{enumerate}
	\item[(1)] $\phi(\mathbf{x})$ is increasing.
	\item[(2)] For level $j$ of component $i$, there exists a vector $(\mathbf{\cdot}_i, \mathbf{x})$ such that
	$\phi({j}_i, \mathbf{x})=j$ while  $\phi({l}_i, \mathbf{x})\ne j$ for $l\ne j$, $i =1, \cdots, n$, and $j = 0, 1, \cdots, M$.
	\item[(3)] $\phi(\mathbf{j})= j$ where $j=0, 1, \ldots, M$.
\end{enumerate}

Condition ($1$) is the monotonicity of the system structure function, which states that if a system component improves its performance, then the system performance may improve or remain unchanged but does not degrade, or vice versa.
Similarly, Condition ($2$) states that each component is \emph{relevant} to the system performance at all of its operational states. Condition ($3$) follows from the above two conditions and describes that if all the components are in the same operational state, say $j$, then the system will also be operating in that state $j$.
Some basic types of multistate coherent systems are as follows:

\begin{enumerate}
	\item[(1)] A series system: $\phi_{series}(\mathbf{x}) = \min\limits_{i=1}^{n} x_i$.
	\item[(2)] A parallel system: $\phi_{parallel}(\mathbf{x}) = \max\limits_{i=1}^{n} x_i$.
	\item[(3)] A $k$-out-of-$n$ structure: $\phi_{k-\mathrm{out-of}-n}(\mathbf{x}) =x_{(n-k+1)}$,  where $x_{(1)}\le x_{(2)}\le \cdots \le x_{(n-1)}\le x_{(n)}$
	is an ascending rearrangement of $x_1, x_2, \cdots, x_{n-1}, x_n$.
\end{enumerate}

In the next sections, we present the HOL formalization of series and parallel multistate coherent systems. We formally verify that they satisfy the essential conditions of multistate coherent systems and also their respective deterministic properties, such as upper and lower bounds, redundancy, and bounds  related to the upper critical connection vector.


\subsection{Multistate Series Structure Function ($\phi_{series}$)} \label{SUBSEC:multistate_series}

A series system can be mathematically expressed as $\phi_{series}(\mathbf{x}) = \min\limits_{i=1}^{n} x_i$. We can formally define it in \holl~as follows:

\begin{definition}
\label{DEF:Phi_series}
\emph{Multistate Series Structure Function} \\{
\textup{\texttt{
			$\vdash_{def}$
			(\kw{series\_MS} [] = (0:num)) $\wedge$ \\ \ \ \ \
  \hspace*{0.8cm}		($\forall$x.\ \kw{series\_MS} [x] = (x:num))  $\wedge$ \\ \ \ \ \
  \hspace*{0.8cm}		($\forall$x xs.\ \kw{series\_MS} ((x:num)::xs) =
  \hspace*{2.0cm}   MIN x (series\_MS xs))
}}}
\end{definition}

The above function takes the states of the components as a list and returns the minimum state by recursively applying the \holl~minimum function \texttt{MIN}. Using Definition~\ref{DEF:Phi_series}, we can formally verify that $\phi_{series}$ satisfies the multistate coherent systems Conditions ($1$-$3$) as:

\begin{theorem}
\label{THM:Phi_series_multi_coherent_sys_cond_1}
\emph{Multistate Coherent Systems Condition $1$ for $\phi_{series}$} \\{
\textup{\texttt{
			$\vdash_{thm}$
			$\forall$(L:(num\#num) list).\ \\
 \hspace*{0.3cm}  \kw{[A1:]} $\neg$NULL L $\wedge$  \\
  \hspace*{0.3cm}  \kw{[A2:]} X\_LOW\_Y (X\_list L) (Y\_list L)  \\
 \hspace*{0.8cm} $\Rightarrow$ (series\_MS (X\_list L) $\le$
   \hspace*{1.6cm} series\_MS (Y\_list L))
}}}
\end{theorem}

The above theorem satisfies the monotonicity property (Condition ($1$)) for the series structure function $\phi_{series}(\mathbf{x})$. Assumption \texttt{A1} states that the components state vectors must not be empty. The functions \texttt{X\_list} and \texttt{Y\_list} are defined in \holl~to extract the component state vectors from a list of pairs, where the first element of the pair are the elements of state vector X and the second element of the pair belongs to the state vector Y, respectively. For example, \texttt{X\_list [(x1,y1),(x2,y2)] = [x1;x2]}. The function \texttt{X\_LOW\_Y} in Assumption \texttt{A2} formally defines the relationship $\mathbf{x} < \mathbf{y}$ explained above and is formalized in \holl~as:

\begin{definition}
\label{DEF:x_low_y}
\emph{Components States in $\mathbf{x}$ are less than Elements in $\mathbf{y}$} \\{
\textup{\texttt{
			$\vdash_{def}$
         (\kw{X\_LOW\_Y} [] = T) $\wedge$ \\ 	
   \hspace*{0.8cm}		($\forall$h t.\ \kw{X\_LOW\_Y} (h::t) =    \\
    \hspace*{1.2cm}  (FST h $\le$ SND h) $\wedge$   \kw{X\_LOW\_Y} t)
}}}
\end{definition}

\noindent where the \holl~functions \texttt{FST} and \texttt{SND} accept a pair and return the first and second element of the pair, respectively. The verification of Theorem~\ref{THM:Phi_series_multi_coherent_sys_cond_1} is mainly based on induction, Definitions~\ref{DEF:Phi_series} and~\ref{DEF:x_low_y} alongwith some arithmetic reasoning.

\begin{theorem}
\label{THM:Phi_series_multi_coherent_sys_cond_2}
\emph{Multistate Coherent Systems Condition $2$ for $\phi_{series}$} \\{
\textup{\texttt{
			$\vdash_{thm}$
			$\forall$l j L i.\ \\
     \hspace*{0.3cm}  \kw{[A1:]}  l$\ \ne$ j $\wedge$   \\
      \hspace*{0.3cm}  \kw{[A2:]}   i\ < LENGTH L $\wedge$ \\
       \hspace*{0.3cm}  \kw{[A3:]}  $\neg$NULL L $\wedge$ \\
      \hspace*{0.3cm}  \kw{[A4:]} ($\forall$x. MEM x L $\Rightarrow$ x > j) $\wedge$ \\
        \hspace*{0.3cm}  \kw{[A5:]} (series\_MS (LUPDATE j i L) = j)    \\
    \hspace*{0.8cm} $\Rightarrow$ (series\_MS (LUPDATE l i L) $\ne$ j)
}}}
\end{theorem}

Assumption \texttt{A1} makes sure that \texttt{l} must not equal to \texttt{j}. Assumption \texttt{A2} ensures that the value of index \texttt{i} must not exceed the length of the list \texttt{L}. Assumption \texttt{A3} is same as Assumption \texttt{A1} of Theorem~\ref{THM:Phi_series_multi_coherent_sys_cond_1}.
Assumption \texttt{A4} guarantees that all the elements in the components state vector \texttt{L} must be greater than State \texttt{j}. Assumption \texttt{A5} ensures that the system acquires the State \texttt{j} only when the component \texttt{i} in the state vector \texttt{L} is updated to the state value \texttt{j} as all other components are in the state greater than \texttt{j}. The \holl~function \texttt{LUPDATE} in above theorem models the notation $(j_i, \mathbf{x})$, i.e., it updates the $i^{th}$ element of vector $\mathbf{x}$ with value \texttt{j}. Theorem~\ref{THM:Phi_series_multi_coherent_sys_cond_2} satisfies the component relevancy property (Condition ($2$)) for the series structure function $\phi_{series}(\mathbf{x})$. The proof process of Theorem~\ref{THM:Phi_series_multi_coherent_sys_cond_2} is based on induction and Definition~\ref{DEF:Phi_series} alongwith some properties of lists.

\begin{theorem}
\label{THM:Phi_series_multi_coherent_sys_cond_3}
\emph{Multistate Coherent Systems Condition $3$ for $\phi_{series}$} \\{
\textup{\texttt{
			$\vdash_{thm}$
          $\forall$L j. \\
    \hspace*{0.3cm}  \kw{[A1:]}  $\neg$NULL L $\wedge$  \\
      \hspace*{0.3cm}  \kw{[A2:]}  ($\forall$x.\ MEM x L $\Rightarrow$  x = j) \\
      \hspace*{0.8cm}   $\Rightarrow$ (series\_MS L = j)
}}}
\end{theorem}

\noindent where Assumption \texttt{A1} in the above theorem is same as Assumption \texttt{A1} of Theorem~\ref{THM:Phi_series_multi_coherent_sys_cond_1}. Assumption \texttt{A2} ensures that all the components must be in State \texttt{j}. This implies that the multistate series structure function $\phi_{series}$ will also be in State \texttt{j} (Condition ($3$)). The verification of Theorem~\ref{THM:Phi_series_multi_coherent_sys_cond_3} is based on Definition~\ref{DEF:Phi_series} and properties of lists alongwith arithmetic reasoning. From Theorems~\ref{THM:Phi_series_multi_coherent_sys_cond_1}-~\ref{THM:Phi_series_multi_coherent_sys_cond_3}, we can conclude that the $\phi_{series}$ is a multistate coherent structure function.


\subsection{Deterministic Properties for $\phi_{series}$} \label{sec:detser}

We now present the HOL formalization of the deterministic properties for $\phi_{series}$, such as upper and lower bounds, system concatenation, and critical upper connection vector bounds, as follows:

\begin{theorem}
\label{THM:Lower_and_upper_bounds_phi_series}
\emph{Lower and Upper Bounds for $\phi_{series}$} \\{
\textup{\texttt{
			$\vdash_{thm}$
			$\forall$X. min\_vec X $\le$ series\_MS X $\wedge$ \\
   \hspace*{1.6cm}    series\_MS X $\le$ max\_vec X	}}}
\end{theorem}

The functions \texttt{min\_vec} and \texttt{max\_vec} take the components state vector \texttt{X} in the form of a list and return the minimum and maximum states, respectively. The above theorem verifies the lower and upper bounds for $\phi_{series}$.

The property $\phi_{series} (\mathbf{x} \vee \mathbf{y}) \ge \phi_{series} (\mathbf{x}) \vee \phi_{series} (\mathbf{y})$ describes the effect of redundancy at the component and system levels for $\phi_{series}$ where the left hand side of the inequality represents the redundancy at the component level and the right hand side gives the redundancy at the system level. We formally verify this inequality in \holl~as follows:

\begin{theorem}
\label{THM:Redundancy_at_comp_sys_level_phi_series}
\emph{Relationship of Redundancy at the Component and System levels} \\{
\textup{\texttt{
			$\vdash_{thm}$
         $\forall$L. series\_MS (X\_OR\_Y L) $\ge$  \\
        \hspace*{1.4cm}   MAX (series\_MS (X\_list L))
          \hspace*{2.6cm}   (series\_MS (Y\_list L))
}}}
\end{theorem}

The function \texttt{X\_OR\_Y} models the notation $\mathbf{x} \vee \mathbf{y}$ where the functions \texttt{X\_list} and \texttt{Y\_list} represent the state vectors $\mathbf{x}$ and $\mathbf{y}$, respectively. The function \texttt{X\_OR\_Y} is formally modeled in \holl~as follows:

\begin{definition}
\label{DEF:function_x_or_y}
\emph{Function $\mathbf{x} \vee \mathbf{y}$} \\{
\textup{\texttt{
			$\vdash_{def}$ (\kw{X\_OR\_Y} [] = []) $\wedge$  \\ 	
   \hspace*{0.7cm}  $(\forall$h t.\  \kw{X\_OR\_Y} (h::t) =   \\
    \hspace*{0.9cm}   (MAX (FST h) (SND h))::(\kw{X\_OR\_Y} t))
}}}
\end{definition}

Similarly, the property $\phi_{series} (\mathbf{x} \wedge \mathbf{y}) = \phi_{series} (\mathbf{x}) \wedge \phi_{series} (\mathbf{y})$ describes that $\phi_{series}$ is compositional in nature, i.e., a system level structure function is composed of component level structure functions. We formally verify this relationship in \holl~as:

\begin{theorem}
\label{THM:Compositional_nature_phi_series}
\emph{Compositional Nature of $\phi_{series}$} \\{
\textup{\texttt{
			$\vdash_{thm}$
         $\forall$L. series\_MS (X\_AND\_Y L) =    \\
        \hspace*{1.4cm}   MIN (series\_MS (X\_list L))  \\
        \hspace*{2.6cm}   (series\_MS (Y\_list L))
}}}
\end{theorem}

The function \texttt{X\_AND\_Y}  models the notation $\mathbf{x} \wedge \mathbf{y}$ and it is formally defined in \holl~as follows:

\begin{definition}
\label{DEF:function_x_and_y}
\emph{Function $\mathbf{x} \wedge \mathbf{y}$} \\{
\textup{\texttt{
			$\vdash_{def}$ (\kw{X\_AND\_Y} [] = []) $\wedge$  \\ 	
   \hspace*{0.6cm}  $(\forall$h t.\  \kw{X\_AND\_Y} (h::t) =   \\
   \hspace*{0.8cm} (MIN (FST h) (SND h))::(\kw{X\_AND\_Y} t))
}}}
\end{definition}

The concept of bounds on $\phi_{series} (\mathbf{x})$ is realized by special kinds of component state vectors, namely $connection$ $vectors$ and $upper$ $critical$ $connection$ $vectors$. They are especially quite useful when the entire system performance is difficult to evaluate and one wants to know the bounds of system performance relative to some known state vectors. We formally define these concepts in \holl~as follows:

\begin{definition}
\label{DEF:connection_vectors_phi_series}
\emph{A vector $\mathbf{x}$ is defined to be a connection vector to level $j$ if $\phi(\mathbf{x}) = j$, $j = 0, 1,  \cdots , M$.} \\{
\textup{\texttt{
			$\vdash_{def}$
            $\forall$X j $\mathtt{\phi}$.\ \kw{con\_vec} X j $\mathtt{\phi}$ =  \\
           \hspace*{3.5cm}  (($\mathtt{\phi}$ X) = (j:num))
}}}
\end{definition}

\begin{definition}
\label{DEF:upper_critical_connection_vectors_phi_series}
\emph{A vector $\mathbf{x}$ is defined to be an upper critical connection vector to level j if $\phi(\mathbf{x}) = j$ and strictly $\mathbf{y}$ $<$ $\mathbf{x}$ implies $\phi(\mathbf{x}) < j$, $j = 0, 1,  \cdots , M$.} \\{
\textup{\texttt{
			$\vdash_{def}$
           $\forall$X j $\phi$.\ \kw{uc\_con\_vec} X j $\phi$ = \\
     \hspace*{0.2cm}  con\_vec X j $\phi$  $\wedge$  \\
   \hspace*{0.2cm} ($\forall$Y.\  (X\_LOWER\_Y ZIP (Y,X)) $\Rightarrow$ $\phi$ Y < j)
}}}
\end{definition}

\noindent where the \holl~function \texttt{ZIP} converts a pair of lists \texttt{$([x_1,. . .,x_n], [y_1,. . .,y_n])$} to a list of pairs \texttt{$[(x_1,y_1),. . .,(x_n,y_n)]$}. The function \texttt{X\_LOWER\_Y} models the strict case of the notation $\mathbf{y} < \mathbf{x}$ similar to the non-strict case of the notation defined by the function \texttt{X\_LOW\_Y} (Definition~\ref{DEF:x_low_y}).

\begin{definition}
\label{DEF:set_upper_critical_connection_vector_level_j}
\emph{Set of Upper Critical Connection Vectors to Level $j$} \\{
\textup{\texttt{
			$\vdash_{def}$
         ($\forall$Y $\phi$ j.\ \kw{uc\_con\_vec\_set0} (0:num)  \\
         \hspace*{1.4cm}  Y j $\phi$ = uc\_con\_vec (Y 0) j $\phi$) $\wedge$  \\
   \hspace*{0.6cm}	($\forall$Y $\phi$ j n.\ \kw{uc\_con\_vec\_set0} \\
   \hspace*{4.0cm} (SUC n) Y j $\phi$ = \\
  \hspace*{1.2cm}	uc\_con\_vec (Y (SUC n)) j $\phi$ $\wedge$  \\
   \hspace*{1.2cm}  \kw{uc\_con\_vec\_set0} n Y j $\phi$)
}}}
\end{definition}

\begin{theorem}
\label{THM:system_is_lower_bounded_by_j_phi_series}
\emph{System is Lower Bounded by Level $j$} \\{
\textup{\texttt{
			$\vdash_{thm}$
         $\forall$n Y j l X. \\
     \hspace*{0.3cm}  \kw{[A1:]}   $\neg$NULL (X) $\wedge$  \\
      \hspace*{0.3cm}  \kw{[A2:]}   n $\ge$ l $\wedge$ \\
    \hspace*{0.3cm}  \kw{[A3:]}    LENGTH (Y l) = LENGTH X  $\wedge$  \\
   \hspace*{0.3cm}  \kw{[A4:]}    X\_LOW\_Y (Y l) X  $\wedge$ \\
        \hspace*{0.3cm}  \kw{[A5:]}  uc\_con\_vec\_set0 n Y j series\_MS     \\
         \hspace*{1.3cm} $\Rightarrow$ (series\_MS X $\ge$ j)
}}}
\end{theorem}

The above property states that if an upper critical vector of a system to level \texttt{j} is known, then the system will be lower bounded by level \texttt{j} for all those state vectors that are upper vectors to the system's upper critical vector.


\subsection{Multistate Parallel Structure Function ($\phi_{parallel}$)}

A parallel system can be expressed mathematically as $\phi_{parallel}(\mathbf{x}) = \max\limits_{i=1}^{n} x_i$.

\begin{definition}
\label{DEF:Phi_parallel}
\emph{Multistate Parallel Structure Function} \\{
\textup{\texttt{
			$\vdash_{def}$
	 (\kw{parallel\_MS} [] = (0:num)) $\wedge$ \\
  \hspace*{0.8cm} ($\forall$x.\ \kw{parallel\_MS} [x] = x:num)  $\wedge$ \\
  \hspace*{0.8cm} ($\forall$x xs.\ \kw{parallel\_MS} ((x:num)::xs) = \\
  \hspace*{2.8cm}  MAX x (\kw{parallel\_MS} xs))
}}}
\end{definition}

The above function is defined similar to Definition~\ref{DEF:Phi_series} except the function \texttt{MAX} is used to model the multistate parallel system. We formally verify Condition $1$ of multistate coherent systems for $\phi_{parallel}$ as the following HOL4 theorem:

\begin{theorem}
\label{THM:Phi_parallel_multi_coherent_sys_cond_1}
\emph{Multistate Coherent Systems Condition $1$ for $\phi_{parallel}$} \\{
\textup{\texttt{
			$\vdash_{thm}$
			$\forall$(L:(num\#num) list).\ \\
   \hspace*{0.3cm}  \kw{[A1:]}    $\neg$NULL L $\wedge$  \\
    \hspace*{0.3cm}  \kw{[A2:]}  X\_LOW\_Y (X\_list L) (Y\_list L)  \\
 \hspace*{1.3cm}  $\Rightarrow$ (parallel\_MS (X\_list L) $\le$ \\
  \hspace*{2.1cm}  parallel\_MS (Y\_list L))
}}}
\end{theorem}

Assumptions \texttt{A1}-\texttt{A2} are the same as that of Theorem~\ref{THM:Phi_series_multi_coherent_sys_cond_1}. The proof process of Theorem~\ref{THM:Phi_parallel_multi_coherent_sys_cond_1} is mainly based on induction and Definition~\ref{DEF:Phi_parallel} alongwith some arithmetic reasoning. Similarly, we formally verify Condition $2$ for $\phi_{parallel}$ as:

\begin{theorem}
\label{THM:Phi_parallel_multi_coherent_sys_cond_2}
\emph{Multistate Coherent Systems Condition $2$ for $\phi_{parallel}$} \\{
\textup{\texttt{
			$\vdash_{thm}$
			$\forall$l j L i.\ \\
    \hspace*{0.1cm}  \kw{[A1:]}  l$\ \ne$ j $\wedge$  \\
     \hspace*{0.1cm}  \kw{[A2:]}  i\ < LENGTH L $\wedge$  \\
      \hspace*{0.1cm}  \kw{[A3:]}   $\neg$NULL L $\wedge$ \\
     \hspace*{0.1cm}  \kw{[A4:]}  ($\forall$x.\  MEM x L $\Rightarrow$ x > j) $\wedge$ \\
   \hspace*{0.1cm}  \kw{[A5:]}   (parallel\_MS (LUPDATE j i L) = j) \\
  \hspace*{0.5cm} $\Rightarrow$ (parallel\_MS (LUPDATE l i L) $\ne$ j)
}}}
\end{theorem}

Assumptions \texttt{A1}-\texttt{A4} are the same as that of Theorem~\ref{THM:Phi_series_multi_coherent_sys_cond_2}. Assumption \texttt{A5} ensures that the
system acquires the State \texttt{j} only when the component \texttt{i} in the state vector \texttt{L} is updated to the state value \texttt{j} as all other components are in the state greater than \texttt{j}. The verification of Theorem~\ref{THM:Phi_parallel_multi_coherent_sys_cond_2} is mainly based on induction, Definition~\ref{DEF:Phi_parallel} alongwith some properties of lists.

\begin{theorem}
\label{THM:Phi_parallel_multi_coherent_sys_cond_3}
\emph{Multistate Coherent Systems Condition $3$ for $\phi_{parallel}$} \\{
\textup{\texttt{
			$\vdash_{thm}$
          $\forall$L j. \\
  \hspace*{0.1cm}  \kw{[A1:]} $\neg$NULL L $\wedge$  \\
   \hspace*{0.1cm}  \kw{[A2:]} ($\forall$x. MEM x L $\Rightarrow$  x = j)  \\
    \hspace*{1.2cm}  $\Rightarrow$ (parallel\_MS L = j)
}}}
\end{theorem}

Assumptions \texttt{A1}-\texttt{A2} are the same as that of Theorem~\ref{THM:Phi_series_multi_coherent_sys_cond_3}. The proof process of Theorem~\ref{THM:Phi_parallel_multi_coherent_sys_cond_3} is mainly based on induction, Definition~\ref{DEF:Phi_parallel} and properties of lists alongwith some arithmetic reasoning.


\subsection{Deterministic Properties of $\phi_{parallel}$}

Similar to Section \ref{sec:detser}, we formally verify the deterministic properties for $\phi_{parallel}$ as follows:

\begin{theorem}
\label{THM:Lower_and_upper_bounds_phi_parallel}
\emph{Lower and Upper Bounds for $\phi_{parallel}$} \\{
\textup{\texttt{
			$\vdash_{thm}$
			$\forall$X. min\_vec X $\le$ parallel\_MS X $\wedge$ \\
   \hspace*{1.6cm}    parallel\_MS X $\le$ max\_vec X	}}}
\end{theorem}

\begin{theorem}
\label{THM:Redundancy_at_comp_sys_level_phi_parallel}
\emph{Relationship of Redundancy at the Component and System levels} \\{
\textup{\texttt{
			$\vdash_{thm}$
         $\forall$L. parallel\_MS (X\_OR\_Y L) =  \\
        \hspace*{1.3cm}   MAX (parallel\_MS (X\_list L)) \\
       \hspace*{2.5cm}  (parallel\_MS (Y\_list L))
}}}
\end{theorem}

It is worth mentioning that the redundancy for $\phi_{parallel}$ at the component level has same effect as the redundancy at the system level which is opposite to the result of Theorem~\ref{THM:Redundancy_at_comp_sys_level_phi_parallel} for $\phi_{series}$, where the component level redundancy surpasses the system level redundancy.

\begin{theorem}
\label{THM:compositional_nature_phi_parallel}
\emph{Compositional Nature of $\phi_{parallel}$} \\{
\textup{\texttt{
			$\vdash_{thm}$
         $\forall$L. parallel\_MS (X\_AND\_Y L) $\le$    \\
        \hspace*{0.8cm}   MIN (parallel\_MS (X\_list L))
        \hspace*{2.0cm}   (parallel\_MS (Y\_list L))
}}}
\end{theorem}

Similarly, Theorem~\ref{THM:compositional_nature_phi_parallel} shows that the sequential concatenation at the component level is worse than the sequential concatenation at the system level for $\phi_{parallel}$. This result is also in contrast to Theorem~\ref{THM:Compositional_nature_phi_series}, which describes that sequential concatenation at the component level has same effect as the sequential concatenation at the system level for $\phi_{series}$.

Similarly, following theorem verifies the bound on $\phi_{parallel}$ described by the upper critical connection vector, similar to the result of Theorem~\ref{THM:system_is_lower_bounded_by_j_phi_series} for $\phi_{series}$.

\begin{theorem}
\label{THM:system_is_lower_bounded_by_j_phi_parallel}
\emph{System is Lower Bounded by Level $j$} \\{
\textup{\texttt{
			$\vdash_{thm}$
         $\forall$n Y j l X. \\
   \hspace*{0.3cm}  \kw{[A1:]} $\neg$NULL (X) $\wedge$ \\
   \hspace*{0.3cm}  \kw{[A2:]} n $\ge$ l $\wedge$ \\
    \hspace*{0.3cm}  \kw{[A3:]}   LENGTH (Y l) = LENGTH X  $\wedge$  \\
  \hspace*{0.3cm}  \kw{[A4:]}   X\_LOW\_Y (Y l) X  $\wedge$ \\
    \hspace*{0.3cm}  \kw{[A5:]} uc\_con\_vec\_set0 n Y j \\
     \hspace*{5.0cm} parallel\_MS  \\
   \hspace*{1.3cm} $\Rightarrow$  (parallel\_MS X $\ge$ j)
}}}
\end{theorem}


\section{Stochastic performance of multistate coherent systems} \label{SEC:stoc_perf_multistate_coherent_systems}

In this section, we formalize the relationship between the probabilistic performance of of multistate coherent systems and their components.
Let $X_i$ be the random state of $i^{th}$ component, with

\begin{equation} \label{eq1}
P[X_i=j]=p_{ij},
\end{equation}

\begin{equation} \label{eq2}
P[X_i\le j]=P_i(j),
\end{equation}

\noindent where $P_i$ be the $performance$ $distribution$ of $i^{th}$ component and  $j=0,1, \cdots, M$ and $i=1,2,\cdots,n$.

\begin{equation} \label{EQ:eq3}
P_i(j)=\sum_{k=0}^{j} p_{ik},
\end{equation}

\begin{equation} \label{EQ:eq4}
P_i(M)=\sum_{k=0}^{M} p_{ik}=1.
\end{equation}

In order to formalize the above equations, we first formally define the events for probability mass and distribution functions in \holl~as:

\begin{definition}
\label{DEF:event_pmf}
\emph{Event for Probability Mass Function} \\{
\textup{\texttt{
			$\vdash_{def}$
			$\forall$X p j. \\
  \hspace*{0.8cm}	\kw{PMF\_random\_state\_event} X p j = \\
  \hspace*{1.2cm}	\{x | (X x) = (j:num)\} $\cap$ p\_space p
}}}
\end{definition}

\begin{definition}
\label{DEF:event_pdf}
\emph{Event for Probability Distribution Function} \\{
\textup{\texttt{
			$\vdash_{def}$
			$\forall$X p j. \\
  \hspace*{0.8cm}	\kw{CDF\_rand\_state\_event} X p j = \\
  \hspace*{1.2cm}	\{x | (X x) $\le$ (j:num)\} $\cap$ p\_space p
}}}
\end{definition}

The function \texttt{PMF\_random\_state\_event} describes the event when a component is acquiring an arbitrary State \texttt{j}. Similarly, the function \texttt{CDF\_rand\_state\_event} models the event for cumulative distribution function (CDF) describing the event when a component is acquiring a state value equal to or less than an arbitrary state value \texttt{j}. Now, the PMF of a random component state can be modeled in
\holl~as follows:

\begin{definition}
\label{DEF:pmf_random_comp_state}
\emph{PMF of a Random Component State} \\{
\textup{\texttt{
			$\vdash_{def}$
			$\forall$X p j. \kw{PMF\_component} X p j = \\
  \hspace*{0.3cm}  prob p (PMF\_random\_state\_event X p j)
}}}
\end{definition}

\noindent where \texttt{X} is the random state variable, \texttt{p} represents the probability space and \texttt{j} represents a state value. The \holl~function \texttt{prob} takes the probability space and an event within the probability space, and returns the probability of the event

The CDF of the random state variable can be modeled in \holl~as follows:

\begin{definition}
\label{DEF:cdf_random_state_variable}
\emph{CDF of the Random State Variable} \\{
\textup{\texttt{
			$\vdash_{def}$
			$\forall$X p j. \kw{CDF\_component} X p j = \\
  \hspace*{0.6cm}	prob p (CDF\_rand\_stat\_event X p j)
}}}
\end{definition}

Using Definitions~\ref{DEF:pmf_random_comp_state} and~\ref{DEF:cdf_random_state_variable}, we can also verify that the performance distribution of a random state variable is equal to sum of its performances of corresponding states as described in Equation~(\ref{EQ:eq3}).

\begin{theorem}
\label{THM:performance_distribution_of_a_random_state_variable}
\emph{Performance Distribution of a Random State Variable} \\{
\textup{\texttt{
			$\vdash_{thm}$
         $\forall$X p j. \\
   \hspace*{0.1cm}  \kw{[A1:]}  prob\_space p $\wedge$ \\
 \hspace*{0.1cm}  \kw{[A2:]}   ($\forall$l.\ PMF\_random\_state\_event X   \\
  \hspace*{5.0cm} p l $\in$ events p) \\
    \hspace*{1.3cm}  $\Rightarrow$  (CDF\_component X p j =   \\
    \hspace*{2.5cm}  $\sum\limits_{k=0}^{j}$ PMF\_component X p k)
}}}
\end{theorem}

Similarly, we formally prove the important property of the performance distribution of component random state variable as described in Equation~(\ref{EQ:eq4}), as follows:

\begin{theorem}
\label{THM:performance_distribution_of_component_random_state_variable}
\emph{Performance Distribution of Component Random State Variable} \\{
\textup{\texttt{
			$\vdash_{thm}$
        $\forall$X p M. \\
    \hspace*{0.1cm}  \kw{[A1:]}  prob\_space p $\wedge$ \\
  \hspace*{0.1cm}  \kw{[A2:]} component\_events\_space X M =  \\
   \hspace*{6.0cm}  p\_space p    \\
    \hspace*{1.3cm} $\Rightarrow$   CDF\_component X p M = 1
}}}
\end{theorem}

\noindent where Assumption \texttt{A1} ensures that \texttt{p} is a valid probability space. Assumption \texttt{A2} asserts that the component state variable has the maximum state value \texttt{M} and all the events, defined by the states of the component, form a valid sample space in \texttt{p}. The function \texttt{component\_events\_space} takes a random component state variable \texttt{X} and its maximum state value \texttt{M}, and returns the union of all the events representing the sample space of \texttt{X}. This function is formally defined in \holl~as follows:

\begin{definition}
\label{DEF:Union_events_rep_sample_space_X}
\emph{Union of all Events Representing the Sample Space of \texttt{X}} \\{
\textup{\texttt{
			$\vdash_{def}$
	  ($\forall$X.\ \kw{component\_events\_space} X 0 =  \\
    \hspace*{4.7cm}	PREIMAGE X {0})  $\wedge$  \\
  \hspace*{0.7cm}	($\forall$X M.\ \kw{component\_events\_space} X  \\
  \hspace*{6.0cm}	 (SUC M) = \\
  \hspace*{1.2cm} \kw{component\_events\_space} X M $\cup$  \\
  \hspace*{3.9cm}	 PREIMAGE X \{SUC M\})
}}}
\end{definition}


\subsection{System Level Performance}

Now, we describe the system level probabilities in \holl. Let $\mathbf{X}=(X_1,\cdots,X_n)$ be the random vector depicting the random state variables of the components $1,\cdots,n$, where the $X_1,\cdots, X_n$ are taken as statistically mutually independent in nature. Then, $\phi(\mathbf{X})$ is the random variable that describes the random state of the multistate coherent system having structure function $\phi(\mathbf{x})$, with

\begin{equation} \label{EQ:eq5}
P[\phi(\mathbf{X})=j]=p_{j}, \ \  \ \ \ j=0,1, \cdots, M.
\end{equation}

\begin{equation} \label{EQ:eq5}
P[\phi(\mathbf{X})\le j]=P(j),\ \ \ \  j=0,1, \cdots, M.
\end{equation}

\noindent where $P$ denotes the \textit{performance distribution} of the system. We formalize the above equations in \holl~as follows:

\begin{definition}
\label{DEF:pmf_phi_state_event}
\emph{PMF of $\phi$ State Event} \\{
\textup{\texttt{
			$\vdash_{def}$
	 $\forall$X $\phi$  p j.\
         \kw{PMF\_PHI\_state\_event} \\
         \hspace*{1.7cm}  ($\phi$:num list -> num) X p j = \\
  \hspace*{0.2cm}	\{x | $\phi$ (MAP ($\lambda$a. a x) X) = (j:num)\} $\cap$ \\
   \hspace*{0.8cm}  p\_space p
}}}
\end{definition}

\begin{definition}
\label{DEF:cdf_phi_state_event}
\emph{CDF of $\phi$ State Event} \\{
\textup{\texttt{
			$\vdash_{def}$
	 $\forall$X $\phi$  p j.\
         \kw{CDF\_PHI\_state\_event}  \\
        \hspace*{1.7cm}  ($\phi$:num list -> num) X p j = \\
  \hspace*{0.2cm}	\{x | $\phi$ (MAP ($\lambda$a. a x) X) $\le$ (j:num)\} $\cap$ \\
  \hspace*{0.8cm}  p\_space p
}}}
\end{definition}

\begin{definition}
\label{DEF:pmf_phi}
\emph{PMF of $\phi$} \\{
\textup{\texttt{
			$\vdash_{def}$
	 $\forall$X $\phi$ X p j.\
     \kw{PMF\_PHI} $\phi$ X p j =  \\
    \hspace*{0.4cm}  prob p (PMF\_PHI\_state\_event $\phi$ X p j)
}}}
\end{definition}

\begin{definition}
\label{DEF:cdf_phi}
\emph{CDF of $\phi$} \\{
\textup{\texttt{
			$\vdash_{def}$
	 $\forall$X $\phi$ X p j.\
      \kw{CDF\_PHI} $\phi$ X p j =  \\
    \hspace*{0.4cm}  prob p (CDF\_PHI\_state\_event $\phi$ X p j)
}}}
\end{definition}

Now, we formalize the probabilistic properties of the system performance, such as its dependency on the performance of its components and its upper and lower bounds in terms of its components performances. We have formally proved these properties for both $\phi_{series}$ and $\phi_{parallel}$ structure functions. If a lower performance component is replaced with a higher performance component, then the system performance must be improved for coherent systems.  We formally verify this property in \holl~as:

\begin{theorem}
\label{THM:dependency_performance_system_component}
\emph{Let $P_i$, $P_i^{'}$ be two possible performance distributions, i = 1, $\cdots$, n. Further assume $P_i(j) \ge P_i^{'}(j)$ for j = 0, 1, $\cdots$, M and i=1, $\cdots$, n. Let P and $P^{'}$ be the corresponding system performance distributions. Then $P(j) \ge P^{'}(j)$ for j = 0, 1, $\cdots$, M.} \\{
\textup{\texttt{
			$\vdash_{thm}$
			$\forall$L j p. \\
\hspace*{0.3cm}  \kw{[A1:]}  $\neg$NULL L $\wedge$  \\
\hspace*{0.3cm}  \kw{[A2:]} prob\_space p $\wedge$ \\
	\hspace*{0.3cm}  \kw{[A3:]} $\forall$l.\ MEM l L   \\
   \hspace*{1.2cm} $\Rightarrow$ CDF\_rand\_stat\_event (FST l)  \\
  \hspace*{4.7cm}  p j $\in$ events p $\wedge$ \\
\hspace*{0.3cm}  \kw{[A4:]} $\forall$l.\ MEM l L   \\
   \hspace*{1.2cm} $\Rightarrow$ CDF\_rand\_stat\_event (SND l)  \\
   \hspace*{4.7cm}  p j $\in$ events p $\wedge$ \\
\hspace*{0.3cm}  \kw{[A5:]} $\forall$l.\ MEM l L  \\
    \hspace*{1.2cm}  $\Rightarrow$ CDF\_rand\_stat\_event (FST l) p \\
\hspace*{0.4cm}   j $\subset$   CDF\_rand\_stat\_event (SND l) p j  \\
	\hspace*{0.2cm} $\Rightarrow$    (CDF\_PHI series\_MS (X\_list L) p j  $\le$  \\
 \hspace*{0.7cm}  CDF\_PHI series\_MS (Y\_list L) p j $\wedge$ \\
\hspace*{0.6cm}  CDF\_PHI parallel\_MS (X\_list L) p j $\le$  \\
 \hspace*{0.7cm}  CDF\_PHI parallel\_MS (Y\_list L) p j)
}}}
\end{theorem}

Assumptions \texttt{A1} and \texttt{A2} are same as Assumptions \texttt{A1} of Theorems~\ref{THM:Phi_series_multi_coherent_sys_cond_1} and~\ref{THM:performance_distribution_of_component_random_state_variable}. Assumptions \texttt{A3} and \texttt{A4} ensure that all the events associated with the component states must belong to the event space \texttt{p}. Assumption \texttt{A5} ensures the condition  $P_i^{'}(j) \le P_i(j)$.

The next property describes the lower bound on the system performance distribution and is formally verified in \holl~as:

\begin{theorem}
\label{THM:lower_bound_on_the_system_performance_distribution}
\emph{Let $P_i$, be the performance distribution of  component i, for i= 1, $\cdots$, n and all the components are mutually independent. Then, for j = 0, 1, $\cdots$, M-1 : $\prod_{i=1}^{n}P_i(j) \le P(j)$.} \\{
\textup{\texttt{
			$\vdash_{thm}$
			$\forall$L j p. \\
\hspace*{0.3cm}  \kw{[A1:]}  $\neg$NULL L $\wedge$  \\
\hspace*{0.3cm}  \kw{[A2:]} prob\_space p $\wedge$ \\
\hspace*{0.3cm}  \kw{[A3:]} ($\forall$l.\ MEM l L  \\
  \hspace*{1.2cm} $\Rightarrow$ (CDF\_rand\_stat\_event l p  \\
    \hspace*{4.7cm}  j $\in$ events p)) $\wedge$ \\
\hspace*{0.3cm}  \kw{[A4:]} (mutual\_indep p (MAP   \\
 \hspace*{0.9cm} ($\lambda$a. CDF\_rand\_stat\_event a p j) L)) \\
\hspace*{0.8cm} $\Rightarrow$  (list\_prod (MAP  \\
\hspace*{1.2cm}  ($\lambda$a. CDF\_component a p j) L) $\le$ \\
 \hspace*{2.3cm} CDF\_PHI series\_MS L p j)
}}}
\end{theorem}

Similarly, we can formally verify the above property for multistate parallel coherent system as follows:

\begin{theorem}
\label{THM:lower_bound_on_the_system_perf_dist_multi_parallel_coherent_systems}
\emph{Theorem~\ref{THM:lower_bound_on_the_system_performance_distribution} for Multistate Parallel Coherent Systems} \\{
\textup{\texttt{
			$\vdash_{thm}$
			$\forall$L j p. \\
\hspace*{0.1cm}  \kw{[A1:]} $\neg$NULL L $\wedge$  \\
\hspace*{0.1cm}  \kw{[A2:]} prob\_space p $\wedge$ \\
\hspace*{0.1cm}  \kw{[A3:]} ($\forall$l.\ MEM l L  \\
 \hspace*{1.0cm} $\Rightarrow$ (CDF\_rand\_stat\_event l p  \\
 \hspace*{4.5cm} j $\in$ events p)) $\wedge$ \\
\hspace*{0.1cm}  \kw{[A4:]}  (mutual\_indep p (MAP  \\
  \hspace*{0.7cm} ($\lambda$a.\ CDF\_rand\_stat\_event a p j) L)) \\
\hspace*{0.6cm} $\Rightarrow$  (list\_prod (MAP   \\
 \hspace*{1.0cm} ($\lambda$a.\ CDF\_component a p j) L) $\le$\\
  \hspace*{2.1cm}  CDF\_PHI parallel\_MS L p j)
}}}
\end{theorem}

The function \texttt{mutual\_indep} adapted from Ahmad et al.'s work~\cite{ahmad2015towards}, defines the mutual independence of the events given in the list. Similarly, the function \texttt{list\_prod} returns the product of all the elements in the given list.

Similarly, the upper bound on the system performance distribution can be formally verified as follows:

\begin{theorem}
\label{THM:lower_bound_on_the_system_perf_dist_multi_parallel_coherent_systems}
\emph{Let $P_i$, be the $i^{th}$ component performance distribution for i= 1, $\cdots$, n and the components are mutually independent. Then, for j = 0, 1, $\cdots$, M-1: $ P(j) \le 1-\prod_{i=1}^{n}\overline{P}_i(j)$.} \\{
\textup{\texttt{
			$\vdash_{thm}$
			$\forall$L j p. \\
\hspace*{0.3cm}  \kw{[A1:]} $\neg$NULL L $\wedge$  \\
\hspace*{0.3cm}  \kw{[A2:]}  prob\_space p $\wedge$ \\
\hspace*{0.3cm}  \kw{[A3:]} ($\forall$l. MEM l L  \\
 \hspace*{1.2cm}  $\Rightarrow$ (CDF\_rand\_stat\_event l p  \\
 \hspace*{4.7cm} j $\in$ events p)) $\wedge$ \\
\hspace*{0.3cm}  \kw{[A4:]} (mutual\_indep p (MAP  \\
   \hspace*{0.5cm} ($\lambda$a.\ CDF\_rand\_stat\_event a p j) L)) \\
\hspace*{0.8cm} $\Rightarrow$   (CDF\_PHI series\_MS L p j $\le$ \\
\hspace*{1.0cm}  1 - list\_prod (MAP  \\
\hspace*{0.7cm} ($\lambda$a.\ compl\_CDF\_component a p j) L))
}}}
\end{theorem}

\noindent where the function \texttt{compl\_CDF\_component} is the compliment function of the function \texttt{CDF\_component}, depicting the notations $\overline{P}_i(j)$ and ${P}_i(j)$, respectively. This function is formally defined in \holl~as follows:

\begin{definition}
\label{DEF:compliment_cdf_comp}
\emph{Compliment Function of \texttt{CDF\_component}} \\{
\textup{\texttt{
			$\vdash_{def}$
  $\forall$ $\phi$ X p.  \\
  \hspace*{0.5cm} \kw{compl\_CDF\_component} X p = \\
   \hspace*{0.8cm} $\lambda$a.\ (1 - CDF\_component X p a)
}}}
\end{definition}

\begin{figure}[!ht]
	\centering
	\includegraphics[width=0.95\linewidth,scale=0.80]{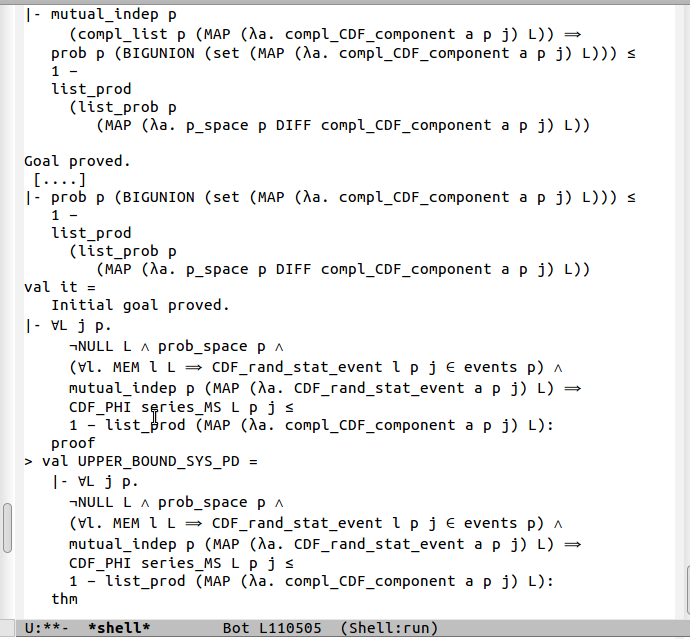}
	\caption{Execution of the Formal Proof of Theorem~\ref{THM:lower_bound_on_the_system_perf_dist_multi_parallel_coherent_systems}}
	\label{fig:hol4_illus_1}
\end{figure}

Figure~\ref{fig:hol4_illus_1} provides an illustration of the execution of the formal proof of Theorem~\ref{THM:lower_bound_on_the_system_perf_dist_multi_parallel_coherent_systems} in HOL$4$

\section{Performance Analysis of Multistate Oil and Gas Pipeline System} \label{SEC:case_study}

\begin{figure}[!ht]
	\centering
	\includegraphics[width=0.7\linewidth,scale=0.35]{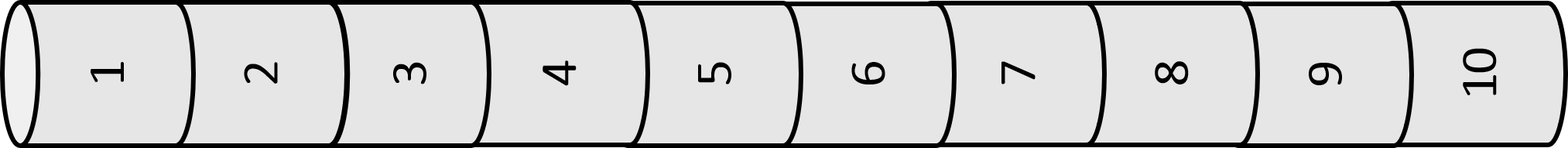}
	\caption{An Oil and Gas Pipeline Depicting $10$ Segments}
	\label{fig:g1}
\end{figure}

The performance analysis of the oil and gas pipelines is necessary to get insights about their life and usability. A typical long-distance oil and gas pipeline can be partitioned into a series connection of $N$ segments, where these segments may be in different operational states based on their individual failure times. The working levels of the pipeline and its segments can acquire any level in $0,\ 1,\ \cdots,\ M$, where $0$ represents the complete failure state and $M$ denotes the perfect working state. For example, a $60$ segment pipeline is analyzed in Zhang et al.'s work~\cite{zhang2008reliability} using the concept of segments but this work considers only two states of operation for each segment. In the following sections, we present the formal performance analysis of a multistate coherent series oil and gas pipeline, as depicted in Figure \ref{fig:g1}. This formal analysis is a two step process. In the first step, we need to formally specify the pipeline segment, which mainly includes the formal modeling of the pipeline segments and their different states probabilities. In the next step, these formal models are used for formally verifying the performance distribution of the series oil and gas pipeline operating in different states.

\subsection{Formal Specifications}

A pipeline segment can be completely specified with its operational states and corresponding state probabilities. We can formally define the segment by specifying its random state variable and the state probabilities as follows:

\begin{definition}
\label{DEF:segment_state_probability}
\emph{Segment State Probability} \\{
\textup{\texttt{
			$\vdash_{def}$
    $\forall$p s. \kw{CDF\_seg} p s =  \\
		\hspace*{0.8cm} (PMF\_component (FST s) p \\
  \hspace*{1.2cm} (FST (SND s)) = (SND (SND s)))
}}}
\end{definition}

\noindent where \texttt{p} is the probability space and \texttt{s} is a tuple $(X_i,j,p_{ij})$ stating that a random state variable $X_i$ is in State $j$ and have probability $p_{ij}$.

The following definition specifies all the pipeline segments.

\begin{definition}
\label{DEF:pipeline_segments}
\emph{Pipeline Segments} \\{
\textup{\texttt{
			$\vdash_{def}$
     $\forall$p S SL. \\
		\hspace*{0.3cm}	\kw{component\_state\_pred} p S SL = \\
	\hspace*{0.4cm}	(CDF\_seg\_list p \\
  \hspace*{0.9cm} (seg\_tuple\_list S SL)) $\wedge$ \\
    	\hspace*{0.4cm}	 (component\_events\_space S  \\
    \hspace*{1.8cm} (LENGTH SL) = p\_space p)
}}}
\end{definition}

\noindent where \texttt{S} represents the random state variable of segment and \texttt{SL} specifies the list of state probability pairs  of the form ${(j,p_{ij})}$ of the random  state variable \texttt{S}. The first predicate in above definition allocates the state probabilities to the corresponding state events using the function \texttt{PMF\_component} while the second condition ensures that the list \texttt{SL} forms a complete event space. The function \texttt{seg\_tuple\_list} takes the random variable \texttt{S} and list of its state probability pairs of the form ${(j,p_{ij})}$ and returns the list of tuples of the form ${(S,j,p_{ij})}$. The function \texttt{CDF\_seg\_list} returns a list of component performances at all operational states obtained from the function \texttt{seg\_tuple\_list} (Definition~\ref{DEF:list_segment_tuples}). These functions are formally defined in \holl~as follows:

\begin{definition}
\label{DEF:list_segment_state_probabilities}
\emph{List of Segment State Probabilities} \\{
\textup{\texttt{
			$\vdash_{def}$
	($\forall$p.\ \kw{CDF\_seg\_list} p [] = T) $\wedge$  \\
	\hspace*{0.8cm}	 ($\forall$p h t.\ \kw{CDF\_seg\_list} p (h::t) = \\
	\hspace*{1.2cm} CDF\_seg p h $\wedge$ \kw{CDF\_seg\_list} p t)
}}}
\end{definition}

\begin{definition}
\label{DEF:list_segment_tuples}
\emph{List of Segment Tuples} \\{
\textup{\texttt{
			$\vdash_{def}$
			$\forall$p. \kw{seg\_tuple\_list} X L =  \\
      \hspace*{2.5cm} MAP ($\lambda$a. (X,a)) L
}}}
\end{definition}

Now, we formally model the complete pipeline, shown in Figure~\ref{fig:g1}, operating as the multistate series coherent system in \holl~as:

\begin{definition}
\label{DEF:performance_distribution_series_pipeline}
\emph{Performance Distribution of Series Pipeline} \\{
\textup{\texttt{
			$\vdash_{def}$
   $\forall$L p j.\ \\
    \hspace*{0.8cm}	\kw{PDF\_pipeline\_series} L p j = \\
    	\hspace*{1.2cm}	 prob p (CDF\_PHI\_state\_event \\
	\hspace*{4.2cm} series\_MS L p j)
}}}
\end{definition}

The above formal specifications of the oil and gas pipeline allow us to reason about the pipeline performance given the performance of its segments.


\subsection{Formal Verification}

Consider a pipeline having $10$ segments, depicted in Figure \ref{fig:g1}, where each segment has states ranging from $0$ to $4$, i.e., there are $5$ operational states for every segment of the pipeline. Let $p_{ij}$ be the performance of the segment $i$ operating in State $j$. Similarly, the notation $P_{ij}$ be the performance distribution of segment $i$ operating in any state $j = 0, 1, \cdots, 4$. Assuming all the segments are performing mutually independent, then the performance distribution of the series pipeline can be mathematically written as:

\begin{equation} \label{eq7}
P_{pipeline}(j) =P[pipeline\le j] = 1-\prod_{i=1}^{10}(1-P_{ij})
\end{equation}

\begin{equation} \label{eq9}
\begin{split}
P_{pipeline}(j) &= 1-\prod\limits_{i=1}^{10}\left(1 - \sum\limits_{k=0}^{j} p_{i,k}\right) \\ &= 1 - ((1-(p_{1,0}+p_{1,1}+\cdots+p_{1,j}))*\\ & \ \ \ (1-(p_{2,0}+p_{2,1}+\cdots+p_{2,j}))*\cdots*\\ &\ \ \ (1-(p_{10,0}+p_{10,1}+\cdots+p_{10,j})))
\end{split}
\end{equation}

Similarly, the performance distribution of the series pipeline operating in the second state, i.e., State $1$ is mathematically expressed as follows:

\begin{equation} \label{eq:per_dis_state_1}
\begin{split}
P_{pipeline}(1) &= 1-\prod\limits_{i=1}^{10}\left(1 - \sum\limits_{k=0}^{1} p_{i,k}\right) \\
  & = 1-\prod\limits_{i=1}^{10}\left(1 - p_{i,0} - p_{i,1}\right) \\
  & = 1 - ((1 - p_{1,0} - p_{1,1}) * (1- p_{2,0} - p_{2,1}) * \\ &
  (1 - p_{3,0} - p_{3,1}) * (1 - p_{4,0} - p_{4,1}) * \\
  & (1 - p_{5,0} - p_{5,1}) * (1 - p_{6,0} - p_{6,1}) * \\
  & (1 - p_{7,0} - p_{7,1}) * (1 - p_{8,0} - p_{8,1}) * \\ &
   (1 - p_{9,0} - p_{9,1}) * (1 - p_{10,0} - p_{10,1}))
\end{split}
\end{equation}

When the multistate system is in the first state, i.e., State $0$, it is in complete failure mode. Therefore, the performance of segment $i$ operating in State $0$ is equal to zero, i.e., $p_{i,0} = 0$ for all $1 \le i \le 10$. Using $p_{i,0} = 0$ in Equation~(\ref{eq:per_dis_state_1}) results in the following equation:

\begin{equation} \label{eq:per_dis_state_1_dead_0}
\begin{split}
P_{pipeline}(1) &= 1-\prod\limits_{i=1}^{10}\left(1 - \sum\limits_{k=0}^{1} p_{i,k}\right) \\
  & = 1-\prod\limits_{i=1}^{10}\left(1 - p_{i,0} - p_{i,1}\right) \\
  & = 1 - ((1 - p_{1,1}) * (1 - p_{2,1}) * (1 - p_{3,1}) * \\ &
   (1 - p_{4,1}) * (1 - p_{5,1}) * (1 - p_{6,1}) * \\
  & (1 - p_{7,1}) * (1 - p_{8,1}) * \\ &
   (1 - p_{9,1}) * (1 - p_{10,1}))
\end{split}
\end{equation}

We formally verify the performance distribution for a series pipeline (Equation~(\ref{eq9})) in \holl~as follows:

\begin{theorem}
\label{THM:performance_distribution_series_pipeline}
\emph{Performance Distribution of Series Pipeline} \\{
\textup{\texttt{
			$\vdash_{thm}$
			$\forall$SL pSL j p.
			\\ \ \   \kw{Let}
			\\ \ \ \ val SL = [S1;S2;...;S10];
			\\ \ \ \ val pSL = [[p1\_0;p1\_1;...;p1\_4];  \\
 \hspace*{2.2cm}  [p2\_0;p2\_1;...;p2\_4];...; \\ \ \ \ \  \ \ \ \ \
 \hspace*{2.2cm}  [p10\_0;p10\_1;...;p10\_4]];
			\\ \ \ \kw{in}
			\\
\hspace*{0.2cm}  \kw{[A1:]} prob\_space p $\wedge$  \\
\hspace*{0.2cm}  \kw{[A2:]}  $\forall$l.\ MEM l (ZIP(SL,pSL)) \\
 \hspace*{2.0cm} $\Rightarrow$ component\_state\_pred \\
 \hspace*{2.8cm} (FST l) p (SND l) $\wedge$  \\
\hspace*{0.2cm}  \kw{[A3:]}  ($\forall$l.\ MEM l SL \\
 \hspace*{2.0cm} $\Rightarrow$ (CDF\_rand\_stat\_event l \\
 \hspace*{3.0cm} p j $\in$ events p))  $\wedge$  \\
\hspace*{0.2cm}  \kw{[A4:]}  mutual\_indep p (MAP ($\lambda$a. \\
\hspace*{0.6cm} CDF\_rand\_stat\_event a p j) SL) \\
\hspace*{0.8cm} $\Rightarrow$   (PDF\_pipeline\_series SL p j = \\
 \hspace*{1.0cm} 1 - ((1 - sum\_list j \\
   \hspace*{2.2cm}  [p1\_0;p1\_1;...;p1\_4])* \\
  \hspace*{1.0cm}  (1 - sum\_list j \\
  \hspace*{2.2cm}  [p2\_0;p2\_1;...;p2\_4])*$\cdots$* \\
    \hspace*{1.0cm} (1 - sum\_list j \\
   \hspace*{2.2cm}   [p10\_0;p10\_1;...;p10\_4]))
}}}
\end{theorem}

\noindent where the list \texttt{SL} specifies the random state variables of all the segments. The list \texttt{pSL} is the list of lists representing the performance of all the pipeline segments operational states. Assumption \texttt{A1} is the same as Assumption \texttt{A1} of Theorem~\ref{THM:performance_distribution_of_component_random_state_variable}. The function \texttt{component\_state\_pred} in Assumption \texttt{A2} gives the specifications of all the segments, i.e., all the random variables in the list \texttt{SL} are assigned to their corresponding performances from the list \texttt{pSL}. Assumptions \texttt{A3} and \texttt{A4} ensure that all the pipeline segments random state variables have their events in the valid probability space \texttt{p} and are statistically mutually independent. Finally, the conclusion of Theorem~\ref{THM:performance_distribution_series_pipeline} models Equation~(\ref{eq9}). The function \texttt{sum\_list} takes a number \texttt{j} and a list of pairs and returns the sum of the second elements of first \texttt{j} pairs.

\begin{definition}
\label{DEF:sum_first_n_elements_list}
\emph{Sum of First $n$ Elements of List} \\{
\textup{\texttt{
			$\vdash_{def}$
			($\forall$n.\ \kw{sum\_list} n [] = []) $\wedge$ \\ \ \ \
	\hspace*{0.8cm} ($\forall$n x xs. \kw{sum\_list} n (x::xs) = \\
  \hspace*{1.3cm} if n = 0 then 0 else  \\
  \hspace*{1.8cm}  (SND x) + \kw{sum\_list} (n - 1) xs)
}}}
\end{definition}

The proof of Theorem~\ref{THM:performance_distribution_series_pipeline} is carried out mainly using Theorems~\ref{THM:performance_distribution_of_a_random_state_variable}, \ref{THM:performance_distribution_of_component_random_state_variable} and \ref{THM:lower_bound_on_the_system_perf_dist_multi_parallel_coherent_systems} and is quite straightforward. It just took about $80$ lines of \holl~code and a few man-hours, which clearly illustrates the effectiveness of our proposed framework. Finally, we formally verify the performance distribution of the series pipeline operating in State $1$ (Equation~(\ref{eq:per_dis_state_1_dead_0})) as follows:

\begin{theorem}
\label{THM:performance_distribution_series_pipeline_state_1}
\emph{Performance Distribution of Series Pipeline Operating in Second state, i.e., State $1$} \\{
\textup{\texttt{
			$\vdash_{thm}$
			$\forall$SL pSL p.
			\\ \ \   \kw{Let}
			\\ \ \ \ val SL = [S1;S2;...;S10];
			\\ \ \ \ val pSL = [[0;p1\_1;...;p1\_4];  \\
 \hspace*{2.2cm}  [0;p2\_1;...;p2\_4];...; \\ \ \ \ \  \ \ \ \ \
 \hspace*{2.2cm}  [0;p10\_1;...;p10\_4]];
			\\ \ \ \kw{in}
			\\
\hspace*{0.2cm}  \kw{[A1:]} prob\_space p $\wedge$  \\
\hspace*{0.2cm}  \kw{[A2:]}  $\forall$l.\ MEM l (ZIP(SL,pSL)) \\
 \hspace*{2.0cm} $\Rightarrow$ component\_state\_pred \\
 \hspace*{2.8cm} (FST l) p (SND l) $\wedge$  \\
\hspace*{0.2cm}  \kw{[A3:]}  ($\forall$l.\ MEM l SL \\
 \hspace*{2.0cm} $\Rightarrow$ (CDF\_rand\_stat\_event l \\
 \hspace*{3.0cm} p 1 $\in$ events p))  $\wedge$  \\
\hspace*{0.2cm}  \kw{[A4:]}  mutual\_indep p (MAP ($\lambda$a. \\
\hspace*{0.6cm} CDF\_rand\_stat\_event a p 1) SL) \\
\hspace*{0.8cm} $\Rightarrow$   (PDF\_pipeline\_series SL p 1 = \\
 \hspace*{1.0cm} 1 - ((1 - p1\_1) * (1 - p2\_1) *  \\
    \hspace*{0.7cm} (1 - p3\_1) * (1 - p4\_1) * \\
    \hspace*{0.7cm} (1 - p5\_1) * (1 - p6\_1) * \\
        \hspace*{0.7cm} (1 - p7\_1) * (1 - p8\_1) * \\
    \hspace*{2.2cm}  (1 - p9\_1) * (1 -  p10\_1))
}}}
\end{theorem}

The verification of the above theorem is mainly based on Theorem~\ref{THM:performance_distribution_series_pipeline_state_1} and is done almost automatically, which clearly illustrate the practical effectiveness of our formally verified Theorem~\ref{THM:performance_distribution_series_pipeline} having generic number of segments and operating states of the multistate system. Thus, the verification mainly involves specializing these generic variables to specific numbers of operating states, such as, $2$ states in Theorem~\ref{THM:performance_distribution_series_pipeline_state_1}.
More details about our formal analysis of the multistate coherent systems can be found at our project webpage~\cite{projectwebperfanaly2021}.

\begin{figure}[!ht]
	\centering
	\includegraphics[width=0.9\linewidth,scale=0.70]{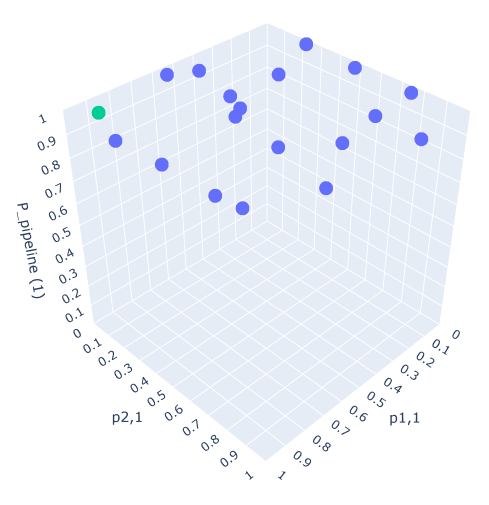}
	\caption{Performance Distribution of Segment $i$ Operating in State $1$ with $p_{i,1} = 0.7$ (Above Average) for $i \ge 3$}
	\label{fig:graph07}
\end{figure}

\begin{figure}[!ht]
	\centering
	\includegraphics[width=0.95\linewidth,scale=0.80]{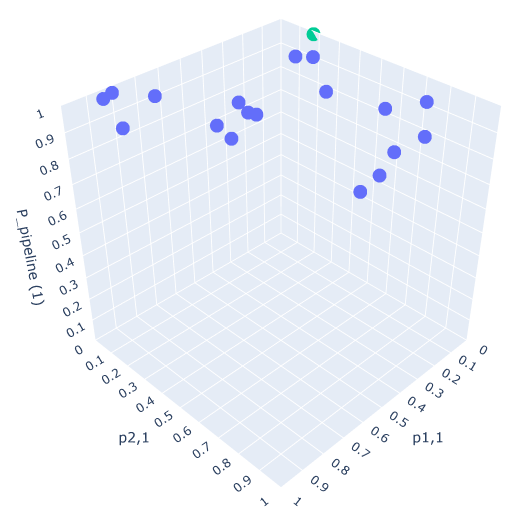}
	\caption{Performance Distribution of Segment $i$ Operating in State $1$ with $p_{i,1} = 0.3$ (Below Average) for $i \ge 3$}
	\label{fig:graph03}
\end{figure}

We encoded our formalized Theorem~\ref{THM:performance_distribution_series_pipeline_state_1} in Python, as given in Algorithm~\ref{Alg:Python_pseudocode}, for different values of $p_{i,1}$, capturing the performance of segment $i$ operating in the second operational state, i.e., State $1$. It provides a graphical representation of the performance distribution of the multistate coherent oil and gas pipeline as shown in Figures~\ref{fig:graph07} and~\ref{fig:graph03}. We set the values of $p_{i,1} = 0.7$ (above average) for $i \ge 3$, whereas, $p_{1,1} \in (0, 1)$ and $p_{2,1} \in (0, 1)$ are represented on $x$ and $y$ axes in Figure~\ref{fig:graph07}. Whereas, the $z$-axis depicts the performance distribution of Segment $i$ operating in State $1$ that has its peak value (green color dot in Figure~\ref{fig:graph07}) at $p_{1,1} = 0.9226$ and $p_{2,1} = 0.1015$. Similarly, Figure~\ref{fig:graph03} captures the performance distribution for a scenario, where $p_{i,1} = 0.3$ (below average) for $i \ge 3$, and $p_{1,1} \in (0, 1)$ and $p_{2,1} \in (0, 1)$. Moreover, the peak of the performance distribution for this scenario exists at $p_{1,1} = 0.00136$ and $p_{2,1} = 0.1698$ and is represented by a green color dot in Figure~\ref{fig:graph03}.

\begin{algorithm}
 \caption{Performance Distribution of Segment \textit{i} operating in State $1$}
     \SetAlgoLined
    \DontPrintSemicolon
\textbf{Input:} performance\_segment, N

\textbf{Output:} Performance indicators: $p_{1,1}$ , $p_{2,1}$ , $P_{pipeline}(1)$

//Initialization

\quad \quad \quad  product\_performance = 1

// Outer Loop for No. of trials

\For  {each, $j \in N$} {

// Inner Loop to compute the product of performance, i.e., \\
\hspace*{0.3cm} $\prod_{i=1}^{10}\left( 1- p_{i , 0} - p_{i , 1} \right)$

 \For {each, $i \in N$}{
           \If {  $\text{performance\_segment} \left[ i \right] <= 1$}{

   product\_performance = product\_performance $\times$ \\
      (1 - performance\_segment $\left[ i \right]$)
}
}

   // Computing performance distribution, \\
     \hspace*{0.8cm}  $P_{pipleline}(1) = 1 - \prod_{i=1}^{10}\left(1- p_{i, 0} - p_{i, 1} \right)$

  $P_{pipeline}(j) = (1 - \text{product\_performance})$

     $p_{1,1} = \text{choose\ probability\ between\ (0 , 1)}$

      $p_{2,1} = \text{choose\ probability\ between\ (0 , 1)}$

        product\_performance = 1

}

 return $([p_{1,1} , p_{2,1} , P_{pipeline}(1)])$

\label{Alg:Python_pseudocode}
\end{algorithm}

The distinguishing features of our proposed HOL formalization for the performance analysis of the multistate coherent systems as compared to the traditional analysis techniques are that it allows us to formally verify generic expressions for multistate oil and gas pipelines, i.e., all of the variables and functions are of generic nature, i.e., universally quantified, presenting any number of segments, any number of operational states and arbitrary probability distributions. Moreover, they can be specialized to perform the formal performance analysis of any real-world multistate coherent system. For example, Theorem~\ref{THM:performance_distribution_series_pipeline} provides the analysis of a multistate oil and gas pipeline system with $10$ number of segments and $5$ operational states. Similarly, Theorem~\ref{THM:performance_distribution_series_pipeline_state_1} presents the analysis of the same system operating in State $1$. Whereas, in the case of simulation based analysis, we need to model and analyze each system individually. Soundness is assured in functions and theorems as every new theorem is verified by applying basic axioms and inference rules or any other previously verified theorems/inference rules. Moreover, this approach guarantees correctness of the analysis by ensuring that all the necessary assumptions for the validity of the result are explicitly present with the theorem. The above mentioned benefits are not assured by any traditional approaches for performing the reliability analysis of the multistate oil and gas pipelines systems that suffer from their inherent limitations, such as, limited computational resources and human error-proneness. Therefore, the reliability analysis presented in this paper undoubtedly depicts the usefulness of the proposed methodology.


\section{Conclusions} \label{SEC:conclusions}

In this paper, we proposed a HOL theorem proving based framework for formally analyzing the reliability of the multistate coherent systems. In particular, we presented the HOL formalization of series and parallel multistate coherent systems. Based on this formalization, we formally verified various deterministic   properties of these systems, such as, upper and lower bounds on structure functions, relationship of redundancy at the component and system levels and the compositional nature of the structure function. Next, we formally verified the probabilistic properties, such as, system level performance. This framework allowed us to formally verify generic mathematical properties about multistate coherent systems with an arbitrary number of components and their states.
Finally, to demonstrate the practical effectiveness of our proposed approach, we performed the formal reliability analysis of the multistate oil and gas pipeline using the HOL4 theorem prover.


\bibliographystyle{sagev}
\bibliography{biblio}

\end{document}